\documentclass[aps,prl,twocolumn,groupedaddress, showpacs]{revtex4-1}
\usepackage{amsmath,amssymb,graphicx}
\usepackage{scrextend}

\begin{document}


\title{Microscopic Theory for the Role of Attractive Forces in the Dynamics of Supercooled Liquids}


\author{Zachary E. Dell}
\affiliation{Department of Physics, University of Illinois, Urbana, IL 61801}

\author{Kenneth S. Schweizer}
\affiliation{Department of Materials Science, University of Illinois, Urbana, IL 61801}
\affiliation{Department of Chemistry, University of Illinois, Urbana, IL 61801}
\affiliation{Frederick Seitz Materials Research Laboratory, University of Illinois, Urbana, IL 61801}


\date{\today}

\begin{abstract}
We formulate a microscopic, no adjustable parameter, theory of activated relaxation in supercooled liquids directly in terms of the repulsive and attractive forces within the framework of pair correlations. Under isochoric conditions, attractive forces can nonperturbatively modify slow dynamics, but at high enough density their influence vanishes. Under isobaric conditions, attractive forces play a minor role. High temperature apparent Arrhenius behavior and density-temperature scaling are predicted. Our results are consistent with recent isochoric simulations and isobaric experiments on a deeply supercooled molecular liquid. The approach can be generalized to treat colloidal gelation and glass melting, and other soft matter slow dynamics problems. 
\end{abstract}

\pacs{64.70.Q-, 64.70.P-, 83.10.Pp}

\maketitle

	The fundamental question of the role of attractive forces in determining the slow dynamics of crowded systems is crucial in diverse soft matter contexts \cite{QRB2001LeckbandIsraelachvili, BR1969Kushner,  JPCM2007Zaccarelli, PR1999BergenholtzFuchs, BouchaudCommentary2010, PRL2009BerthierTarjus, PR2010BerthierTarjus, JCP2011BerthierTarjus, EPJ2011BerthierTarjus, JPC2014Dyre, PRL2010PedersenSchroder,  JPCM2013BohlingVeldhorst}.  Strong, short range attractions can trigger aggregation, gelation and emergent elasticity in colloidal, protein and macromolecular systems  \cite{QRB2001LeckbandIsraelachvili, BR1969Kushner,  JPCM2007Zaccarelli, PR1999BergenholtzFuchs}. The role of slowly varying attractive forces in supercooled liquid dynamics and glass formation is also a critical open question \cite{BouchaudCommentary2010, PRL2009BerthierTarjus, PR2010BerthierTarjus, JCP2011BerthierTarjus, EPJ2011BerthierTarjus, JPC2014Dyre, PRL2010PedersenSchroder,  JPCM2013BohlingVeldhorst}. For all these systems, the construction of a predictive microscopic theory that accurately incorporates attractive forces remains a major challenge. In this Letter we formulate a new statistical dynamical approach broadly relevant to these problems. For concreteness, and because of its fundamental interest, we focus on supercooled liquids. 
	
 Given the van der Waals (vdW) idea that the equilibrium structure of non-associated liquids is dominated by the repulsive branch of the interparticle potential \cite{JCP1971WeeksChandler,S1967Widom,1986HansenMcDonald}, one might expect repulsions dominate slow dynamics. However, recent constant volume simulations \cite{PRL2009BerthierTarjus, PR2010BerthierTarjus, JCP2011BerthierTarjus, EPJ2011BerthierTarjus} of binary sphere mixtures, which probe the initial $\sim$5 orders of magnitude of slowing down, have challenged this idea. They found that the Lenard-Jones (LJ) liquid and its Weeks-Chandler-Andersen (WCA) analog, that contains only the repulsive branch of the potential, indeed exhibit nearly identical equilibrium structure, but at lower liquid-like densities and temperatures the attractive forces slow down relaxation in a non-perturbative manner \cite{PRL2009BerthierTarjus, PR2010BerthierTarjus, JCP2011BerthierTarjus, EPJ2011BerthierTarjus}. Key findings include the following \cite{PRL2009BerthierTarjus, PR2010BerthierTarjus, JCP2011BerthierTarjus, EPJ2011BerthierTarjus}. (i) The large dynamical differences between the LJ and WCA liquids decrease, and ultimately vanish, as the fluid density is significantly increased. (ii) At relatively high temperatures, an apparent Arrhenius behavior is found for both systems over roughly one decade in time with a barrier that grows as a power law with density. (iii) LJ liquid relaxation times at different densities collapse by scaling temperature with the high temperature activation barrier, but such a collapse fails for the WCA fluid. (iv) The ``onset'' temperature at which apparent Arrhenius behavior begins to fail scales with the Arrhenius barrier height \cite{JCP2011BerthierTarjus}.
	
The above simulation findings have been argued \cite{BouchaudCommentary2010, EPJ2011BerthierTarjus} to contradict all existing force-level ``microscopic'' theories (e.g., mode coupling theory (MCT) \cite{2008Gotze,PR1987KirkpatrickWolynes}, nonlinear Langevin theory (NLE) \cite{JCP2005Schweizer}), and thus pose a major open problem in glass physics. It was suggested \cite{EPJ2011BerthierTarjus} that the origin of this failure might be their neglect of higher order than pair correlations. Subsequent simulations found temperature-dependent triplet static correlations do differ for LJ and WCA fluids \cite{JNS2006CasaliniMcGrath, RPP2005RolandHensel-Bielowka}. Moreover, the ``point-to-set'' equilibrium length scale (determined by beyond pair correlation function information) correlates well with the dynamical differences of the two fluids \cite{JCP2013Coslovich}.  
	
In this Letter we re-formulate the starting point for constructing microscopic dynamic theories to explicitly treat attractive forces at the simplest pair correlation level. The key new idea is to analyze the slowly relaxing component of the force-force time correlation function associated with caging directly in terms of the bare forces in real space. This avoids replacing Newtonian forces by effective potentials determined solely by pair structure, a ubiquitous approximation \cite{1986HansenMcDonald, 2008Gotze,PR1987KirkpatrickWolynes,JCP2005Schweizer} that results in theories that are effectively ``blind'' to the dynamical differences between WCA and LJ liquids \cite{PR2010BerthierTarjus, EPJ2011BerthierTarjus}. The predictions of our approach are in good agreement with isochoric simulations \cite{PRL2009BerthierTarjus, PR2010BerthierTarjus, JCP2011BerthierTarjus, EPJ2011BerthierTarjus, PRL2010PedersenSchroder} and isobaric experiments on molecular liquids \cite{PRL2012HockyMarkland,JCP2013BudzienHeffernan, JNS1994RosslerWarschewske}.  
	
	The foundation, or starting point, for many microscopic dynamical theories is the force-force time correlation function, $K(t) = \left< \vec{f}_0(0) \cdot \vec{f}_0(t) \right> $, where $\vec{f}_0(t)$ is the total force on a tagged spherical particle due to its surroundings \cite{2008Gotze, PR1987KirkpatrickWolynes, JCP2005Schweizer, 2001Zwanzig}. Its calculation involves the full many body dynamics and thus a closure approximation must be formulated. In the ideal MCT and single particle na\"{i}ve MCT (NMCT) \cite{2008Gotze, PR1987KirkpatrickWolynes}, the standard closure projects real forces onto the slow bilinear density mode, and four point correlations are factorized into products of pair correlations in a Gaussian manner, which in Fourier space yields:
\begin{align}
	K(t) &= \frac{\beta \rho}{3} \int \frac{d\vec{k}}{\left(2\pi\right)^3} \;\left| \vec{M}(k) \right|^2 S(k) \; \Gamma_s(k,t) \; \Gamma_c(k,t) ,
	\label{eqn:1}
\end{align}
where $ \beta = 1/k_BT$  is the inverse thermal energy, $\rho$ is the fluid number density,  $S(k) = 1 + \rho h(k)$ is the static structure factor,   $h(r) = g(r) -1$ is the nonrandom part of the pair correlation function $g(r)$, and  $\Gamma_s (\Gamma_c )$ is the single particle (collective) dynamic structure factor normalized to unity at $t=0$. Real forces are replaced by an effective force vertex  $\vec{M}(k)$ in Eq. (\ref{eqn:1}) determined entirely by $g(r)$ or $S(k)$\cite{PR1987KirkpatrickWolynes}: 
\begin{align}
	\vec{M}_{NMCT}(k) &= k\,C(k)\,\hat{k},
	\label{eqn:2}
\end{align}
where the direct correlation function is \linebreak$C(k) = \rho^{-1} \left[ 1-S^{-1}(k)\right]$  and the real space effective force is  $k_BT \vec{\nabla} C(r)$. Use of the projection idea implies the dramatic dynamical differences of dense WCA and LJ fluids found in the simulations cannot be captured.
	
To explicitly include the bare forces we re-formulate the dynamical vertex of NMCT based on alternative idea we call the Projectionless Dynamics Theory (PDT). Inspiration comes from prior work in chemical and polymer physics in the normal liquid regime  \cite{JCP1982SchweizerChandler, JCP1989Schweizer1, 1965RiceGray}. Technical details are in the supplementary material (SM) \footnote{\label{foot:SM}See Supplemental Material [below], which includes Refs.  \cite{JCP2014MirigianSchweizer2, JCP2006SaltzmanSchweizer, PRL2009BrambillaEl-Masri, PRL2002PuertasFuchs,NM2002Sciortino, DellPhan}}, but the essential idea is to first analyze the force-force time correlation function in real space as:
\begin{align}
\hspace{-0.1 in}
	K(t) &= \frac{\beta}{3} \int d\vec{r} \int d\vec{r}\,'  \vec{f}(r) \cdot \vec{f}(r') \left< \rho_2 \left(\vec{r}, 0\right) \rho_2\left(\vec{r}\,', t\right)\right> \nonumber \\
	&= \frac{\beta}{3} \int d\vec{r} \int d\vec{r}\,'  \vec{f}(r) \cdot \vec{f}(r') \rho^2 g(r) g(r') \Gamma(\vec{r}, \vec{r}\,', t), 
	\label{eqn:3}
\end{align}
where $\vec{f}(r) = - \vec{\nabla}u(r)$ is the interparticle force (where $r$ is now a field variable), $\rho_2 (\vec{r},t)$ is the instantaneous fluid density a distance $\vec{r}$  from a tagged particle at the origin, at time $t$, and $\left< \rho_2(\vec{r},t)\right> = \rho g(r)$. The object $\Gamma =  \left< \Delta \rho_2 \left(\vec{r}, 0\right) \Delta \rho_2\left(\vec{r}\,', t\right)\right>  / \left( \left< \rho_2(r)\right>\left< \rho_2(r')\right> \right)$ is a multi-point space-time correlation of fluid collective density fluctuations \textit{in the vicinity} of the tagged particle \textit{relative to} the average density inhomogeneity, where \linebreak $\Delta \rho_2\left(\vec{r}, t\right)= \rho_2\left(\vec{r}, t\right) - \rho g(r)$ ; it is approximated by its bulk liquid form factorized to the pair correlation level \cite{JCP1982SchweizerChandler, JCP1989Schweizer1}. The resulting $K(t)$  then has exactly the same form as Eq. (\ref{eqn:1}) but with a different force vertex given by
\begin{align}
	\vec{M}_{PDT}(k) &= \int d\vec{r} g(r) \vec{f}(r) e^{-i \vec{k} \cdot \vec{r}},
	\label{eqn:4}
\end{align}
which is a Fourier-resolved structurally-averaged Newtonian force. The qualitatively new feature is that the real forces now directly enter, and thus identical equilibrium pair structure does \textit{not} imply identical dynamics. 	
  
 \begin{figure}[b!]
	\hspace{-0.25in}
 	\includegraphics[width=0.51\textwidth]{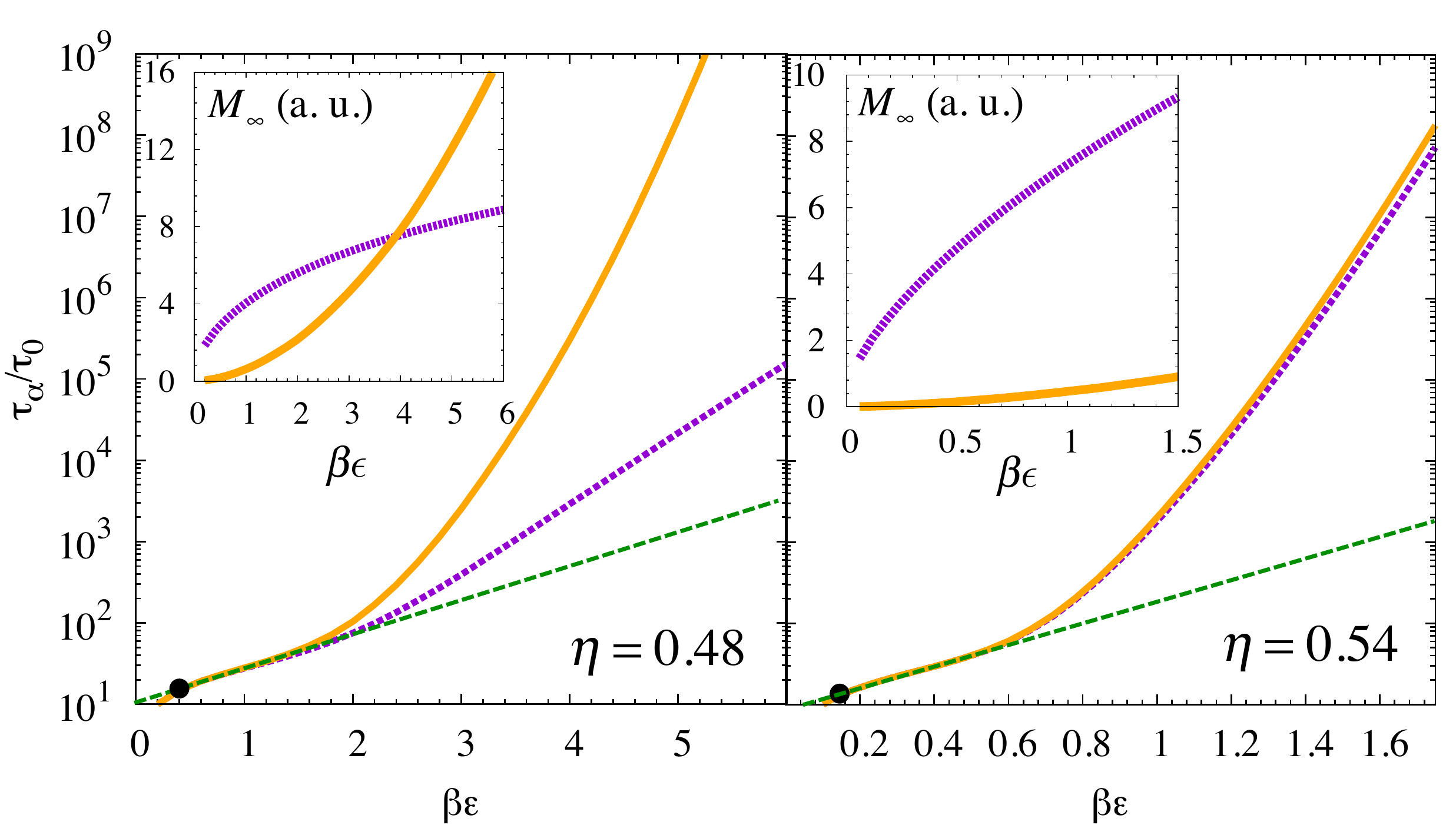}%
 	\caption{\label{fig:1} Non-dimensionalized alpha relaxation times for the LJ (orange, solid) and WCA (purple, dashed) fluids at two packing fractions as a function of dimensionless inverse temperature. For thermal systems, $\tau_0 \equiv (24\rho \sigma^2)^{-1} \sqrt{M/\pi k_BT}$, where $M$ is the particle mass  \cite{JCP2014MirigianSchweizer1}. The black points denote the predicted emergence of a barrier (ideal NMCT crossover), while the green dashed line shows the high temperature Arrhenius behavior. (Inset) The average effective attractive (orange, solid) and repulsive (purple, dashed) contributions to the force vertex, in arbitrary units, for the same packing fractions.}
 \end{figure}	         
The slow dynamics experimentally probed in the deeply supercooled regime, and also the precursor regime accessible to simulation, involves activated motion \cite{PRL2009BerthierTarjus, PR2010BerthierTarjus, JCP2011BerthierTarjus, EPJ2011BerthierTarjus, PR2009Cavagna, RMP2011BerthierBiroli}. Thus, to implement the PDT idea requires a theory of activated relaxation formulated at the level of forces. We employ the well-tested ``Elastically Collective Nonlinear Langevin Equation'' (ECNLE) theory \cite{JCP2014MirigianSchweizer1,JPCL2013MirigianSchweizer}. Based on using the NMCT force vertex, this approach has been shown to accurately capture alpha relaxation in hard sphere fluids and colloidal suspensions over 5-6 decades \cite{JCP2014MirigianSchweizer1}, and molecular liquids over 14 decades based on adopting a lightly coarse-grained mapping to an effective hard sphere fluid \cite{JPCL2013MirigianSchweizer}. Relevant technical details are reviewed in the SM [29]. Briefly, the key physical idea is that knowledge of the slowly decaying component in time of the force memory function in Eq. (\ref{eqn:1}), when combined with the local equilibrium approximation that two particles move relative to each other in a manner that preserves their spatial correlation as determined by $g(r)$, allows for the self-consistent construction of the effective force a single particle experiences due to its local environment as a function of its instantaneous scalar displacement, $r$. This effective force is written as the gradient of a (defined) ``dynamic free energy'', $- \partial F_{dyn}(r)/\partial r$, which enters a stochastic NLE for the tagged particle trajectory. Integration of this force yields $F_{dyn}(r)$. Longer range collective effects enter via the cooperative elastic distortion of the surrounding fluid required to accommodate the irreversible, large amplitude local hopping event described by  $F_{dyn}(r)$ \cite{JCP2014MirigianSchweizer1,JPCL2013MirigianSchweizer}.  The alpha relaxation event has a mixed local-nonlocal character, with a total barrier determined by coupled cage and elastic contributions computed from the dynamic free energy. The alpha time is identified as the mean barrier hopping time computed \cite{1965RiceGray, PR2009Cavagna, RMP2011BerthierBiroli} using Kramers theory \cite{2001Zwanzig}. Crucially, in the PDT framework the basic structure of the ECNLE approach remains \textit{unchanged}, but the fundamental starting point is now Eqs. (\ref{eqn:1}) and (\ref{eqn:4}), not (\ref{eqn:1}) and (\ref{eqn:2}). Thus, \textit{both} pair structure and bare forces influence all aspects of the theory.

 \begin{figure}[]
	\hspace{-0.25in}
 	\includegraphics[width=0.51\textwidth]{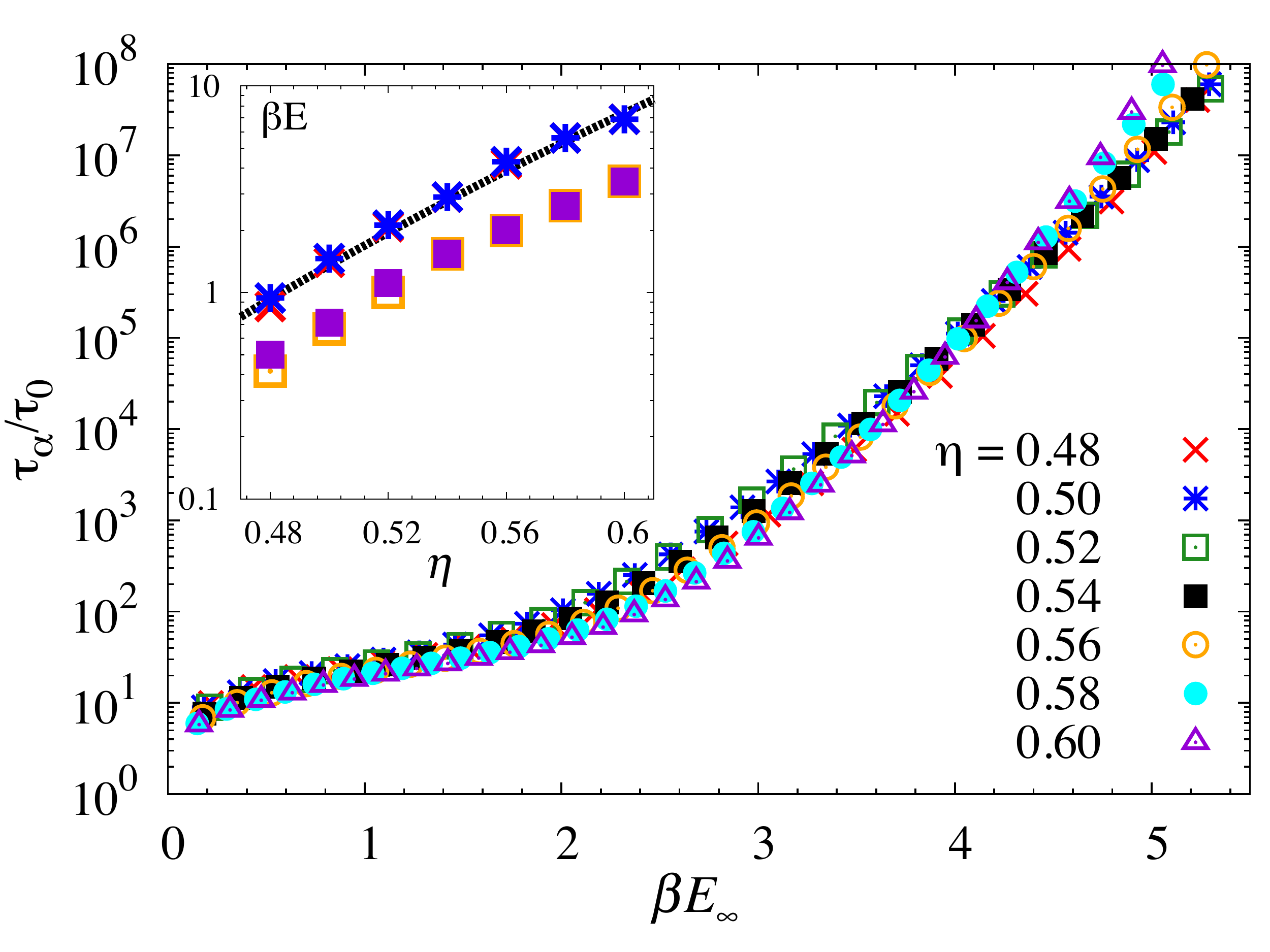}%
 	\caption{\label{fig:2} Collapse of the non-dimensionalized alpha times for the isochoric LJ systems at different packing fractions. Temperature is scaled by the apparent Arrhenius barrier, $E_{\infty}(\eta)$. (Inset) $\beta E_{\infty}$ for LJ (blue, stars) and WCA (red, crosses) fluids (almost indistinguishable), compared to the onset temperature $k_bT_{on}$ (LJ, purple, closed squares; WCA, orange, open squares). The black dashed line is the power law $\beta E_\infty \propto \eta^{9.3}$.}
	\vspace{-0.15in}
 \end{figure}	
 
	We first compare PDT theory predictions for the hard sphere fluid to its analog based on Eq. (\ref{eqn:2}). We find that the NMCT and PDT force vertices for the local $kd>2\pi$ regime are analytically \textit{identical} for dense fluids,  $M(k) \propto g(d) \cos(kd) /(kd)$  \cite{JCP2007SchweizerYatsenko}. The full numerical treatment reveals that both theories predict qualitatively identical density-dependent alpha relaxation times. Quantitatively, use of the NMCT force vertex yields results that agree better with experiment and simulation (see SM) [29]. 

For thermal liquids with attractive interactions, we propose a hybrid approach, in analogy with prior successful microscopic theories of diverse dynamical phenomena that treat the repulsive and slowly varying attractive forces differently  \cite{JCP1982SchweizerChandler,JCP1989Schweizer1,1965RiceGray}. Specifically, we adopt the NMCT vertex for repulsive forces and the PDT vertex for attractive forces: 
\begin{align}
	\left |\vec{M}(k) \right|^2 &= k^2C^2(k) + \left| \int d\vec{r} \; g(r) \;\vec{f}_{att}(r) e^{-i \vec{k} \cdot \vec{r}}\right|^2,
	\label{eqn:5}
\end{align}
where $\vec{f}_{att}$  is the attractive part of the LJ force. For the WCA fluid, only the first term is present. For LJ liquids, the cross term in Eq. (\ref{eqn:5}) is dropped for multiple reasons. (a) It is the simplest (seemingly inevitable) approximation consistent with the use of different dynamic closures for repulsive and attractive forces. (b) Physically, one expects cross correlations are weak since for vdW liquids the attractive and repulsive forces vary on different length scales. (c) The PDT approximation for $\Gamma(\vec{r} - \vec{r}\;', t)$ is known to be more accurate for slowly varying attractions than harsh repulsions \cite{JNS1994RosslerWarschewske}.  

To implement the theory, the WCA repulsion is mapped to an effective hard sphere using the Barker-Henderson (BH) \cite{1986HansenMcDonald,JCP1967BarkerHenderson} expression \linebreak $d_{eff} = \int_0^{2^{1/6}} dr \left[ 1-e^{-\beta U_{WCA}(r)}\right] $ . This mapping is reliable based on recent simulations \cite{JCP2013Coslovich}. Fluid structure is computed using Percus-Yevick (PY) theory \cite{1986HansenMcDonald} with a temperature-dependent effective packing fraction, $\eta_{eff}(T) = \left(d_{eff}(\beta\epsilon)/ \sigma\right)^3 \eta$, where  $\eta = \pi \rho \sigma^3/6$, and $\epsilon$ and $\sigma$ are the LJ energy and length scale, respectively. To isolate the dynamical consequences of attractive forces, the literal vdW picture that $g(r)$ of the LJ and WCA liquids are identical is adopted \cite{JCP1971WeeksChandler,S1967Widom,1986HansenMcDonald}. While the BH mapping and PY theory become less accurate at high densities, no qualitative changes to our results are expected if alternative approximations are employed. Moreover, neither accurate integral equation theory nor simulation data for the WCA $g(r)$ of a \textit{one}-component liquid in the (deeply) supercooled regime are available. Most importantly, the essential leading order origin of our new results is \textit{not} related to pair structure, but rather the explicit accounting for attractive forces on slow dynamics.

  \begin{figure}[]
	\hspace{-0.25in}
 	\includegraphics[width=0.51\textwidth]{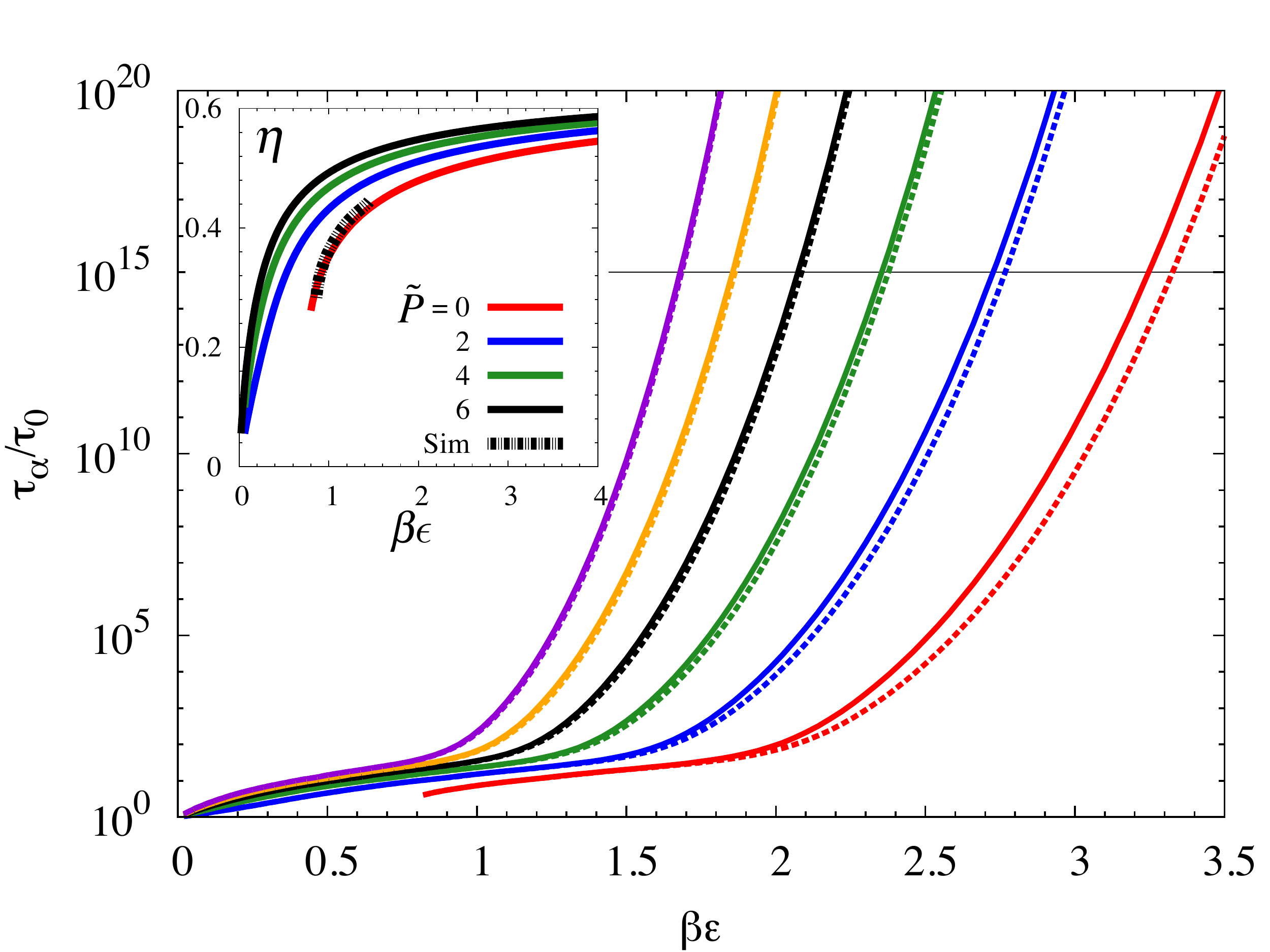}%
 	\caption{\label{fig:3} Dimensionless mean alpha times for LJ (solid) and WCA (dashed) fluids as function of scaled inverse temperature at reduced pressures (right to left) of  $\tilde{P} = 0,\,2,\,4,\,6,\,8,\,10$. The horizontal line illustrates the kinetic vitrification based on $\tau_\alpha(T_g) = 100\,s$ and $\tau_0 = 0.1\,ps$. (Inset) Model equation of state results (curves; see SM) for $\tilde{P} = 0,\,2,\,4,\,6$  (right to left). The black hashed curve shows the fit to simulation data \cite{HT2003BoltachevBaidakov} of the one-component LJ fluid and should be compared to the $\tilde{P} = 0$ (red) curve. }
	\vspace{-0.18in}
 \end{figure}
Under isochoric conditions, $\rho$ and $\eta$ are fixed, but $\eta_{eff}$ grows with cooling via $d_{eff}(\beta\epsilon)$. Representative calculations are shown in Figure \ref{fig:1} for $\eta = 0.48$ and $\eta = 0.54$. For $\eta = 0.48$, the LJ fluid relaxes much slower than its WCA fluid analog at lower temperatures. As $\eta$ increases, these differences smoothly decrease (not shown), and the relaxation times of the two systems are nearly identical at $\eta = 0.54$. These results are in accord with the simulation trends \cite{PRL2009BerthierTarjus, PR2010BerthierTarjus, JCP2011BerthierTarjus, EPJ2011BerthierTarjus}. To develop an intuitive understanding, we compute the long wavelength ($k=0$ in Eq. (\ref{eqn:4})) effective forces that enter the vertex: $M_{\infty, R} \equiv 4 \pi k_BT d_{eff}^2 g(d_{eff})$  for repulsions and $M_{\infty, A} \equiv \int d\vec{r} g(r) f_{att}(r)$ for attractions. The inset of Fig. \ref{fig:1} shows that for $\eta = 0.48$ the repulsive forces dominate at high temperatures where the LJ and WCA relaxation times are similar. The attractive force contribution grows faster than the repulsive analog with cooling and ultimately dominates, consistent with the main frame results. For $\eta = 0.54$ the repulsions dominate at all temperatures.
 
	As seen in simulation \cite{PRL2009BerthierTarjus, PR2010BerthierTarjus, JCP2011BerthierTarjus, EPJ2011BerthierTarjus}, Figure \ref{fig:1} shows that an apparent Arrhenius behavior is predicted at high temperature which is physically due to the unimportance of the collective elasticity aspect of the alpha relaxation process. One can ask whether the theoretical relaxation times for different packing fractions collapse if temperature is scaled by the apparent Arrhenius barrier, $E_{\infty}(\eta)$. In agreement with simulations \cite{PRL2009BerthierTarjus,JCP2011BerthierTarjus}, for WCA fluids no collapse is found (see SM [29]), but for LJ fluids Fig. 2 shows an excellent collapse over 7 decades. The inset shows the Arrhenius barriers are nearly identical for both fluids, and grow as $\beta E_{\infty} \propto \eta^{9.3}$. The high apparent power law exponent (simulation  \cite{PRL2009BerthierTarjus} finds $\sim$5) is expected if the continuous repulsion is replaced by an effective hard sphere potential \cite{JPC2014Dyre}; our exponent value is in excellent agreement with simulations that explored consequences of the WCA to hard sphere mapping  \cite{JCP2013BudzienHeffernan}. We have also computed an ``onset temperature'', $T_{on}$, defined as when the apparent Arrhenius behavior first fails. From Fig. 2 we find $E_{\infty} \approx 2 k_B T_{on}$, consistent with simulation \cite{PRL2009BerthierTarjus, PR2010BerthierTarjus, JCP2011BerthierTarjus, EPJ2011BerthierTarjus}. All the theoretical results discussed above are in good agreement with the trends found in the isochoric simulations performed in the dynamic precursor regime \cite{PRL2009BerthierTarjus, PR2010BerthierTarjus, JCP2011BerthierTarjus, EPJ2011BerthierTarjus}.

\begin{figure}[b!]
	\vspace{-0.3in}
	\hspace{-0.25in}
 	\includegraphics[width=0.51\textwidth]{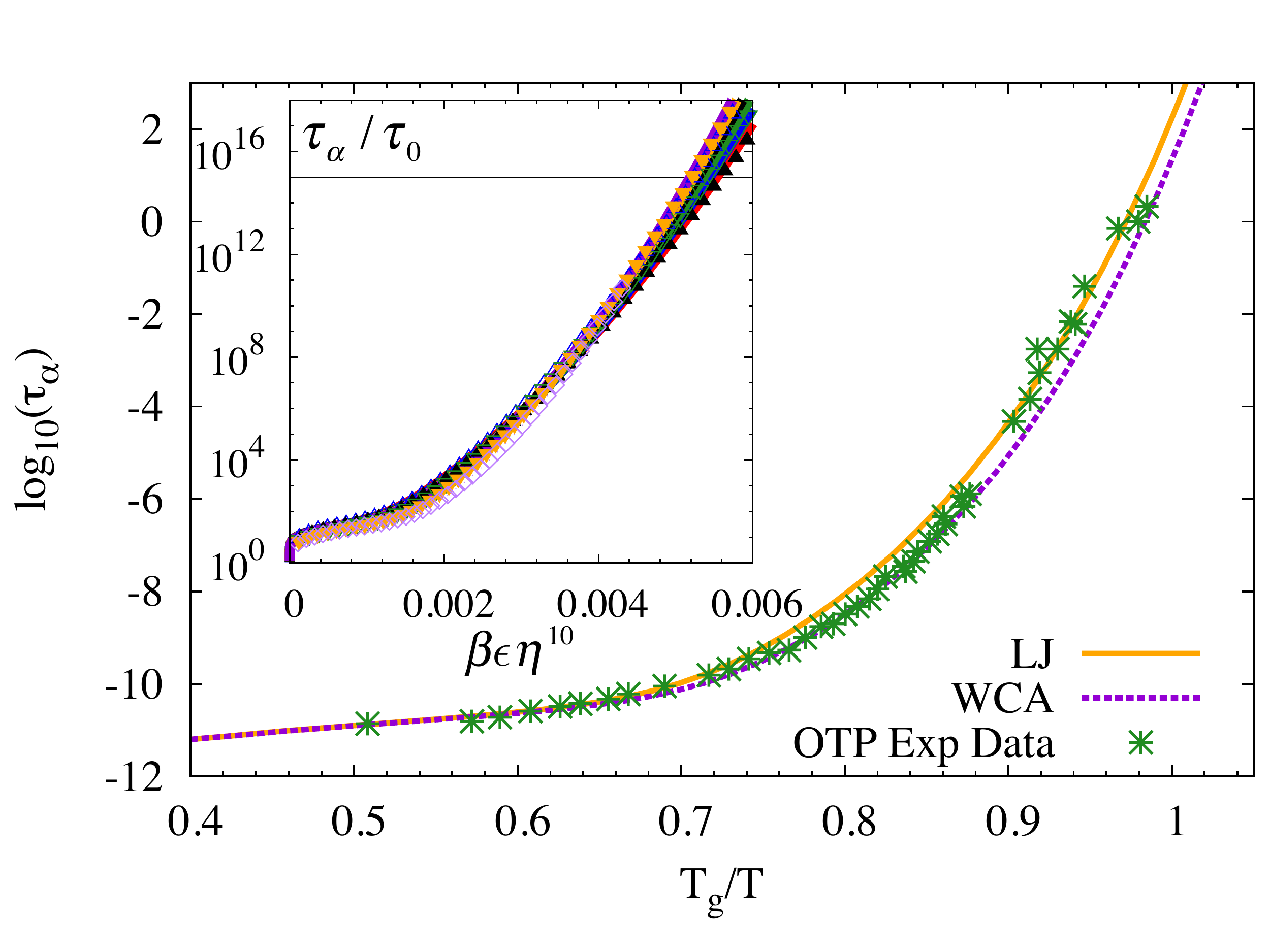}%
 	\caption{\label{fig:4} Logarithm of the mean alpha time (in seconds) versus reduced inverse temperature for the LJ (orange, solid) and WCA (purple, dashed) fluids at $\tilde{P} = 0$, compared to experimental OTP data (green stars) \cite{JNS1994RosslerWarschewske}. The theory curves are shifted vertically to match the high temperature experimental relaxation times.  (Inset) Collapse of the dimensionless alpha times for $\tilde{P} = 0,\,2,\,4,\,6,\,8,\,10$ (curves) and isochoric   $\eta = 0.50,\,0.52,\,0.54,\,0.56,\,0.58 $ (points) conditions with the reduced variable $\beta \epsilon \eta^{10}$. The horizontal line has the same meaning as in Figure 3. }
	 \end{figure}
	Isochoric simulations have also shown that a system interacting via a repulsive inverse power law (IPL) potential, $u_{IPL}(r) = A \epsilon (\sigma/r)^n$, has the same $g(r)$ as the LJ fluid if $A$ and $\epsilon$ are properly tuned \cite{PRL2010PedersenSchroder}. The relaxation times of the LJ and IPL fluids are then found to be nearly identical \cite{PRL2010PedersenSchroder}. In the SM [29] we show that our theory is consistent with this ``hidden scale invariance'' feature and the idea that the dynamical differences between the LJ and WCA fluids is repulsive force truncation \cite{JPC2014Dyre, PRL2010PedersenSchroder}.  

	We now consider experimental systems, which are typically studied at constant pressure and over 14 or more decades in relaxation time \cite{PRL2012HockyMarkland, PR2009Cavagna, RMP2011BerthierBiroli}. We employ a model LJ equation-of-state \cite{HT2003BoltachevBaidakov} (see SM [29]) to perform constant reduced pressure ($\tilde{P} \equiv \beta P \sigma^3$) calculations. The effective packing fraction of the reference hard sphere fluid now varies with temperature due to both an increase of effective particle size $d_{eff}$  with cooling and thermal contraction ($\eta$ increases). Results for the \textit{dynamically} LJ and WCA fluids (with the \textit{same} structural input) are shown in Figure 3. The two fluids have nearly identical relaxation times. At atmospheric pressure ($\tilde{P} = 0$), a one-decade difference is visible, which vanishes as pressure increases because density grows with cooling (Fig. 3 inset). 

	Quantitative contact with isobaric experiments is made based on Fig. 3. A kinetic vitrification temperature $T_g$ is defined as when $\tau_\alpha(T_g) \equiv 10^{15} \tau_0 \simeq 100\,s$ for a typical $\tau_0 \simeq 0.1\,ps$  (horizontal line in Fig. 3). For LJ liquids at atmospheric pressure we find $k_BT_g =0.31\epsilon$, and a fragility of $m_{P = 1 \, atm} = 62$ significantly larger than its isochoric analog of $m_V \approx 26$. This fragility difference is consistent with experiment \cite{PRL2012HockyMarkland}. For the LJ liquid, the theory also properly predicts $T_g$ increases and fragility decreases with pressure (not shown). The vdW liquid orthoterphenyl (OTP) has roughly $\epsilon/k_B \approx 700\, K$ \footnote{$\epsilon_{OTP}$ is roughly estimated from the known value for benzene ($\epsilon_{benz}/k_B = 377 \, K$  \cite{ME1996CuadrosCachadina}) by using the vdW idea that it is proportional to the boiling temperature $T_{boil} \propto \epsilon$, and the ratio $T_{B, OTP}/T_{B, benz} \approx 2$} \cite{ME1996CuadrosCachadina}. Using this, we obtain $T_g = 216 \,K$, in reasonable accord with the experimental $T_g = 246\, K$ \cite{PRL2012HockyMarkland, JNS1994RosslerWarschewske}. Figure 4 demonstrates that the full relaxation time profiles in the reduced inverse temperature Angell representation (vertically shifted to match the high temperature OTP Arrhenius data \cite{JNS1994RosslerWarschewske}) are in excellent agreement with experiment. 

	The inset of Fig. 4 attempts to collapse \textit{both} the isobaric and isochoric LJ liquid relaxation times over a wide range of densities and pressures. The result is consistent with density-temperature scaling \cite{JPC2014Dyre,PRL2012HockyMarkland}. The inset also shows that the density scaling exponent is high ($\sim$10), consistent with recent simulations that mapped WCA repulsions to effective hard spheres \cite{JCP2013BudzienHeffernan} and as expected based on isomorph theory \cite{JPC2014Dyre, PRL2010PedersenSchroder,JPCM2013BohlingVeldhorst}.

	 In conclusion, a new approach for constructing microscopic force-based theories of slow dynamics that explicitly includes attractive forces has been developed at the level of pair correlations. Under isochoric conditions, the attractive forces can have a major effect on supercooled liquid dynamics but as density increases their influence vanishes. Under isobaric conditions, attractive forces are much less important due to thermal contraction. Our results are consistent with recent simulations \cite{PRL2009BerthierTarjus, PR2010BerthierTarjus, JCP2011BerthierTarjus, EPJ2011BerthierTarjus} and experiments \cite{PRL2012HockyMarkland, JNS1994RosslerWarschewske}. The theoretical approach can be applied to more complex soft matter systems. For example, colloidal gels where strong and short range attractive forces induce transient bonding \cite{JPCM2007Zaccarelli} which is explicitly described at the force level using PDT. Although beyond the scope of this Letter, we do find that the essential features of the ``re-entrant glass melting'' phenomenon induced by a short range attraction \cite{JPCM2007Zaccarelli, S2002PhamPuertas,JPC2005ReichmanRabani, JCP2006KaufmanWeitz} is captured by the PDT-ECNLE approach, as briefly discussed in the SM [29]. More generally, the new force vertex idea can be employed in the dynamic free energy framework previously applied to study activated dynamics in glass and gel forming materials composed of nonspherical colloids \cite{PR2011TripathySchweizer,JCP2009TripathySchweizer,PR2009ZhangSchweizer,PR2011ZhangSchweizer}, polymers \cite{M2015MirigianSchweizer} and soft repulsive colloids \cite{EL2010YangSchweizer, JCP2011YangSchweizer}.  

\begin{acknowledgments}
This work was supported by DOE-BES under Grant No. DE-FG02-07ER46471 administered through the Frederick Seitz Materials Research Laboratory. Helpful discussions with Ryan Jadrich and Anh Phan are gratefully acknowledged. We thank John McCoy for bringing Ref. [22] to our attention.
\end{acknowledgments}

\bibliography{/Users/Flory/Documents/BibDesk-Papers/Refs}

\widetext
\clearpage

\begin{center}
\textbf{\large Supplemental Materials: Microscopic Theory for the Role of Attractive Forces in 
the Dynamics of Supercooled Liquids}
\end{center}
\setcounter{equation}{0}
\setcounter{figure}{0}
\setcounter{table}{0}
\setcounter{page}{1}
\makeatletter
\renewcommand{\theequation}{S\arabic{equation}}
\renewcommand{\thefigure}{S\arabic{figure}}

Here we present additional technical details and results concerning seven topics: (i) derivation of the projectionless dynamics theory (PDT), (ii) review of the basics of the elastically cooperative nonlinear Langevin equation (ECNLE) theory \cite{JPCL2013MirigianSchweizer, JCP2014MirigianSchweizer1, JCP2014MirigianSchweizer2} required to perform the calculations in the main text, (iii) relaxation time calculations for the hard sphere fluid and comparison to experiment and simulation, (iv) demonstration of the non-collapse of the predicted WCA fluid relaxation times as a function of temperature for various packing fractions, (v) demonstration that PDT theory predicts that an inverse power law (IPL) repulsive potential can reproduce the behavior of the isochoric LJ liquid, (vi) demonstration that PDT theory predicts ``re-entrant glass melting'' effects in attractive colloidal suspensions ,and (vii) model equation-of-state employed for isobaric calculations.

\section{A. Projectionless Dynamic Theory}
\label{sec:SA}
	For any microscopic theory of single particle motion the key quantity is the force-force time correlation function associated with a tagged particle, $K(t) = \left< \vec{f}_0(0) \cdot \vec{f}_0(t) \right> $ \cite{2008Gotze,PR1987KirkpatrickWolynes,JCP2005Schweizer,JCP2006SaltzmanSchweizer,2001Zwanzig}. Instead of the usual MCT projection of the forces onto the slow bilinear density mode, in PDT the real forces are retained and $K(t)$ is first exactly written in real space in terms of a specific two-body density \cite{JCP1989Schweizer1,JCP1982SchweizerChandler} as:
\begin{align}
K(t) \equiv \left< \vec{f}_0(0) \cdot \vec{f}_0(t) \right> &= \frac{\beta}{3} \int d\vec{r} \int d\vec{r}\,'  \vec{f}(r) \cdot \vec{f}(r') \left< \rho_2 \left(\vec{r}, 0\right) \rho_2\left(\vec{r}\,', t\right)\right> \nonumber \\
	&= \frac{\beta}{3} \int d\vec{r} \int d\vec{r}\,'  \vec{f}(r) \cdot \vec{f}(r') \rho^2 g(r) g(r') \Gamma(\vec{r}, \vec{r}\,', t), 
\label{eqn:SM1}
\end{align}
Here, $\vec{f}(r) = - \vec{\nabla}u(r)$ is the interparticle force (where $r $is now a field variable), $\rho_2 (\vec{r},t)$ is the instantaneous matrix particle density a distance $\vec{r}$ from the tagged particle at time $t$, $\rho g(r)$ is its ensemble average, and $\Gamma =  \left< \Delta \rho_2 \left(\vec{r}, 0\right) \Delta \rho_2\left(\vec{r}\,', t\right)\right>  / \left( \left< \rho_2(r)\right>\left< \rho_2(r')\right> \right)$ is a type of \textit{conditional} multi-point time correlation function where $\Delta \rho_2\left(\vec{r}, t\right)= \rho_2\left(\vec{r}, t\right) - \rho g(r)$. The complex object $\Gamma$ describes the space-time correlation of matrix density fluctuations \textit{in the vicinity} of the tagged particle \textit{relative} to the average density inhomogeneity, $\rho g(r)$.

The key to making progress is to invoke a real space factorization scheme to close the theory at the level of two body correlations. Specifically, we adopt \cite{JCP1989Schweizer1,JCP1982SchweizerChandler} 
\begin{align}
\Gamma(\vec{r} -\vec{r}\,', t) \approx \int d\vec{R} \;\Gamma_s(\vec{R},t)\,S(\vec{r} -\vec{r}\,'+\vec{R},t)
\label{eqn:SM2}
\end{align}
where $\Gamma_s$ and $S$ are the self and collective Van Hove function respectively \cite{1986HansenMcDonald}, which reflect the two parallel channels for force relaxation. Physically, Eq. (\ref{eqn:SM2}) can be viewed as replacing the required multi-point object by its factorized form in the bulk liquid. This approximation has been \textit{a priori} argued to be best when relatively longer wavelength force fluctuations are more important in Eq. (\ref{eqn:SM1}) [6, 7]. We note that at $t=0$, Eq. (\ref{eqn:SM2}) reduces to $\left< \Delta \rho_2 (r) \Delta \rho_2(r')\right> \approx \rho g(r) g(r') S(|\vec{r} - \vec{r} \,'|)$  which corresponds to the classic Kirkwood superposition approximation for 3-body static correlations in liquids \cite{1986HansenMcDonald}.   	

  	Using Eq. (\ref{eqn:SM2}) in Eq. (\ref{eqn:SM1}), performing a Fourier transform, and taking the long time limit to obtain the arrested (at the na\"{i}ve MCT level) part of $K(t)$, one obtains:
\begin{align}
K(t\rightarrow \infty) 	&= \frac{\beta \rho}{3} \int \frac{d\vec{k}}{(2\pi)^3}  \left| \vec{M}_{PDT}(k) \right|^2 S(k) \; \Gamma_s(k,t\rightarrow \infty)\Gamma_c(k,t\rightarrow \infty) \nonumber \\
&=  \frac{\beta \rho}{3} \int \frac{d\vec{k}}{(2\pi)^3}  \left| \vec{M}_{PDT}(k) \right|^2 S(k) \; e^{-k^2 r_L^2(1+S^{-1}(k))/6}\label{eqn:SM3}
\end{align}
where $\Gamma_i(k, t\rightarrow \infty)$ are the self and collective Debye-Waller factors which are explicitly expressed in the usual NMCT Gaussian form  \cite{JCP2005Schweizer, JCP2006SaltzmanSchweizer,JCP2007SchweizerYatsenko} in the second line. Eq. (\ref{eqn:SM3}) is identical to the na\"{i}ve MCT force-force correlation function  \cite{JCP2005Schweizer, JCP2006SaltzmanSchweizer} except the effective force vertex is:
\begin{align}
	\vec{M}_{PDT}(k) &= \int d\vec{r} g(r) \vec{f}(r) e^{-i \vec{k} \cdot \vec{r}},
	\label{eqn:SM4}
\end{align}
instead of $\vec{M}_{NMCT}(k) = k\,C(k)\,\hat{k}$.

\section{B. Calculation of the Alpha Relaxation Time in ECNLE Theory}
We implement the PDT idea in the context of the presently most advanced particle and force level predictive microscopic approach of single particle activated relaxation, the ``Elastically Collective Nonlinear Langevin Equation'' (ECNLE) theory \cite{JPCL2013MirigianSchweizer, JCP2014MirigianSchweizer1, JCP2014MirigianSchweizer2}. This approach includes, in a no adjustable parameter manner, coupled large amplitude, cage scale hopping motion and the long range cooperative elastic distortion of particles outside the cage region required to allow the local re-arrangement to occur. The starting point is the nonlinear Langevin equation (NLE)  which stochastically describes  \cite{JCP2005Schweizer, JCP2006SaltzmanSchweizer}  the scalar displacement, $r(t)$, of a tagged sphere (diameter, $d$): $\zeta_s dr/dt = - \partial F_{dyn}(r)/ \partial r + \xi(t)$  , where $\zeta_s$ is the known short time friction constant \cite{JPCL2013MirigianSchweizer, JCP2014MirigianSchweizer1, JCP2014MirigianSchweizer2} and $\xi(t)$ the corresponding white noise random force. The key object is the dynamic free energy, the gradient of which self-consistently determines the force on a moving particle due to its surroundings. Its general form is:
\begin{align}
\beta F_{dyn}(r, \eta) &= \frac{3}{2} \ln \left( \frac{3 d^2}{2r^2} \right) - \frac{3 \eta}{\pi^3 d^3} \int^\infty_0 dk \left| \vec{M}(k) \right|^2 \frac{S(k)}{1+S^{-1}(k)} e^{-k^2 r^2(1+S^{-1}(k))/6}
\label{eqn:SM5}
\end{align}
where $\beta = 1/k_BT$ and $\eta = \rho \pi d^3/6$ is the fluid packing fraction. The second term captures caging effects via the effective force vertex, $\vec{M}(k)$, from either NMCT or PDT ideas. Equation (\ref{eqn:SM5}) with the NMCT vertex is the centerpiece of the prior stochastic NLE theory which captures \textit{local} uncooperative activated hopping  \cite{JPCL2013MirigianSchweizer, JCP2014MirigianSchweizer1, JCP2014MirigianSchweizer2, JCP2005Schweizer, JCP2006SaltzmanSchweizer} .

	The alpha relaxation time in ECNLE theory is \cite{JPCL2013MirigianSchweizer, JCP2014MirigianSchweizer1, JCP2014MirigianSchweizer2} $\tau_\alpha = \tau_s + \tau_{hop}$. Here, $\tau_s$ describes short time relaxation in the absence of barriers and involves only binary collisions with non-self-consistent local cage corrections \cite{JPCL2013MirigianSchweizer, JCP2014MirigianSchweizer1, JCP2014MirigianSchweizer2,1986HansenMcDonald}, and $\tau_{hop}$ is associated with the activated hopping process due to cage rearrangement and elastic distortion \cite{JPCL2013MirigianSchweizer, JCP2014MirigianSchweizer1, JCP2014MirigianSchweizer2}. The local cage contribution to $\tau_{hop}$, defined as $\tau_{NLE}$, follows from a Kramers calculation of the mean first passage time \cite{2001Zwanzig} to cross the barrier based on the NLE dynamic free energy, $F_{dyn}(r)$ \cite{JCP2005Schweizer, JCP2006SaltzmanSchweizer}:
\begin{align}
\tau_{NLE} &= \frac{2\tau_s}{d^2} \int_{r_L}^{r_B} dr \;e^{\beta F_{dyn}(r)} \;\int_0^r dr' \; e^{-\beta F_{dyn}(r')}
\label{eqn:SM6} 
\end{align}
The outer integral varies from the minimum of the dynamic free energy $r_L$ (Òlocalization lengthÓ) to the barrier position $r_B$. For barriers modestly higher than thermal energy, Eq. (\ref{eqn:SM6}) reduces to the more standard version of Kramer's theory \cite{2001Zwanzig}:
\begin{align}
\frac{\tau_{NLE}}{\tau_s} &= \frac{2\pi}{\sqrt{K_L K_B}} e^{\beta F_B}
\label{eqn:SM7} 
\end{align}
where $K_i$ is the absolute value of the dimensionless curvature of $F_{dyn}(r)$ (in units of $k_BT/d^2$) at its local minimum or maximum, and $F_B$ is the barrier. 
	
The collective elastic barrier is associated with the long range, harmonic, spontaneous fluctuations of particles outside the cage region that is required to allow the large amplitude hopping event to occur, thereby yielding a total barrier of $F_{tot} = F_{B} + F_{elastic}$ \cite{JPCL2013MirigianSchweizer, JCP2014MirigianSchweizer1, JCP2014MirigianSchweizer2}. The hopping time is then:
\begin{align}
\tau_{hop} \equiv \tau_{NLE} \,e^{\beta F_{elastic}} \approx \tau_s \frac{2\pi}{\sqrt{K_L K_B}} e^{\beta (F_B+F_{elastic})}
\label{eqn:SM8} 
\end{align}
where the second approximate equality is applicable for a barrier modestly larger than thermal energy. The elastic barrier is explicitly \cite{JPCL2013MirigianSchweizer, JCP2014MirigianSchweizer1, JCP2014MirigianSchweizer2}:
\begin{align}
\beta F_{elastic} &\approx 12 \eta \left( \Delta r_{eff} \right)^2 \frac{r_{cage}}{d^3} K_L
\label{eqn:SM9} 
\end{align}
where $r_{cage}$ is the location of the first minimum of $g(r)$, $\Delta r_{eff} \simeq 3(\Delta r)^2/(32r_{cage})$ and $\Delta r = r_B - r_L$.

Equations (\ref{eqn:SM5})-(\ref{eqn:SM9}) allow calculation of the mean alpha relaxation time with no adjustable parameters \cite{JPCL2013MirigianSchweizer, JCP2014MirigianSchweizer1, JCP2014MirigianSchweizer2}. Note that for small barriers, $\tau_\alpha \rightarrow \tau_s$ smoothly since $\tau_{NLE} \rightarrow 0$ in Eq. (\ref{eqn:SM6}) continuously as $F_B \rightarrow 0$ and $\Delta r \rightarrow 0$. Hence, the relaxation time smoothly crosses over from its high temperature non-activated form to its activated low temperature form. 

\section{C. Predictions for the Hard Sphere Fluid}
	For hard sphere fluids, at first sight the PDT approach may seem undefined since the force $\vec{f}(r)$ does not exist. The latter fact is sometimes invoked to motivate the projection approximation in the MCT approach to ``shield'' the singular potential \cite{1986HansenMcDonald} and render it a non-infinite effective or pseudo potential (direct correlation function). However, using the continuity and non-singular nature of the cavity distribution function, $y(r) = e^{\beta U_{HS}(r)} g(r)$, and the identity \cite{1986HansenMcDonald}  $g(r) \vec{f}(r) = k_BT \, g(r) \, \delta(r-d^+) \, \hat{r}$, a simple analytic form for Eq. (\ref{eqn:SM4}) can be easily derived:
\begin{align}
	\vec{M}_{HS, PDT}(k) &= 4\pi d^2 k_BT \, g(d) \, j_1(kd) \, \hat{k}
	\label{eqn:SM10}
\end{align}
where $g(d)$ is the contact value of the pair distribution function, and $j_1(x) = \sin x/x^2 - \cos x/ x$.  

 \begin{figure}[t!]
	\vspace{-0.25in}
 	\includegraphics[width=0.6\textwidth]{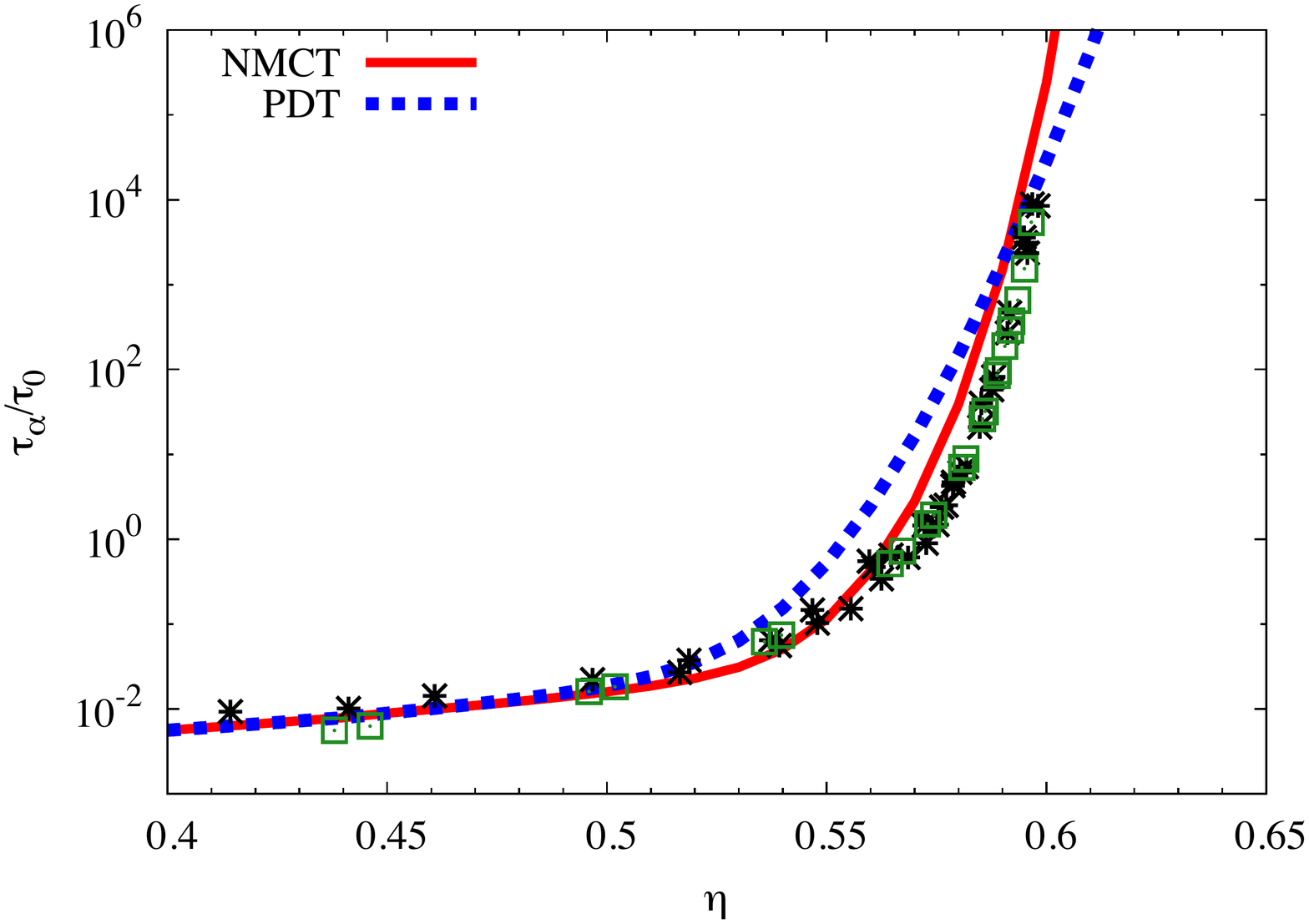}%
	\vspace{-0.5in}
 	\caption{\label{fig:SM1}  ECNLE theory calculation of the hard sphere fluid mean alpha relaxation time non-dimensionalized by the bare diffusion time \cite{JPCL2013MirigianSchweizer, JCP2014MirigianSchweizer1, JCP2014MirigianSchweizer2}  $\tau_0 \equiv d^2/D_0$, where $D_0$ is the bare diffusion constant, as a function of packing fraction  . Results are shown based on  using the projected (red, solid) and projectionless (blue, dashed) effective force vertices. Simulation (green squares) and colloid experimental (black stars) results \cite{PRL2009BrambillaEl-Masri} are shown for comparison. }
	\vspace{-0.1in}
 \end{figure}	

To qualitatively compare the force vertices in the NMCT and PDT approaches, one can analyze the important high wavevector limit, $kd>>1$ (so-called Òultra-localÓ limit of NLE theory \cite{JCP2007SchweizerYatsenko}). In this limit  \cite{1986HansenMcDonald} $C(k) \rightarrow 4\pi k^{-2} \,g(d) \,d\cos(kd)$ , and thus the NMCT vertex becomes $kC(k) \rightarrow  4\pi d \,g(d) \cos(kd)/k$. This result is \textit{identical} to Eq. (\ref{eqn:SM10}) at high wavevectors, strongly suggesting the basic physics captured by the NMCT or PDT based approaches is very similar for hard spheres.  

	To directly compare the predictions of ECNLE theory using either the projected or projectionless force vertex we numerically compute the mean alpha relaxation times in Figure \ref{fig:SM1}. The results are qualitatively similar in all respects. Quantitatively, there are differences. By carefully analyzing the results we find the difference is primarily due to the predicted packing fraction dependence of the jump distance that enters the elastic barrier in Eq. (\ref{eqn:SM9}). We employed the NMCT force vertex for the repulsive forces in our analysis of the WCA and LJ fluids in the main text. Beyond its virtue of being more quantitatively accurate, this practical strategy is internally consistent in that the statistical mechanical approximation employed to construct the PDT (Eq. (\ref{eqn:SM2})) has been a priori argued to be valid for forces that vary relatively slowly in space, which is not true for the hard core potential but applicable to the attractive branch of the LJ potential  \cite{JCP1989Schweizer1,JCP1982SchweizerChandler}. This buttresses our adoption of the hybrid approach of Eq. (\ref{eqn:5}) in the main text from a theoretical perspective.

\section{D. Non-Collapse of Relaxation Times in the WCA Fluid}
 \begin{figure}[h!]
 	\includegraphics[width=0.6\textwidth]{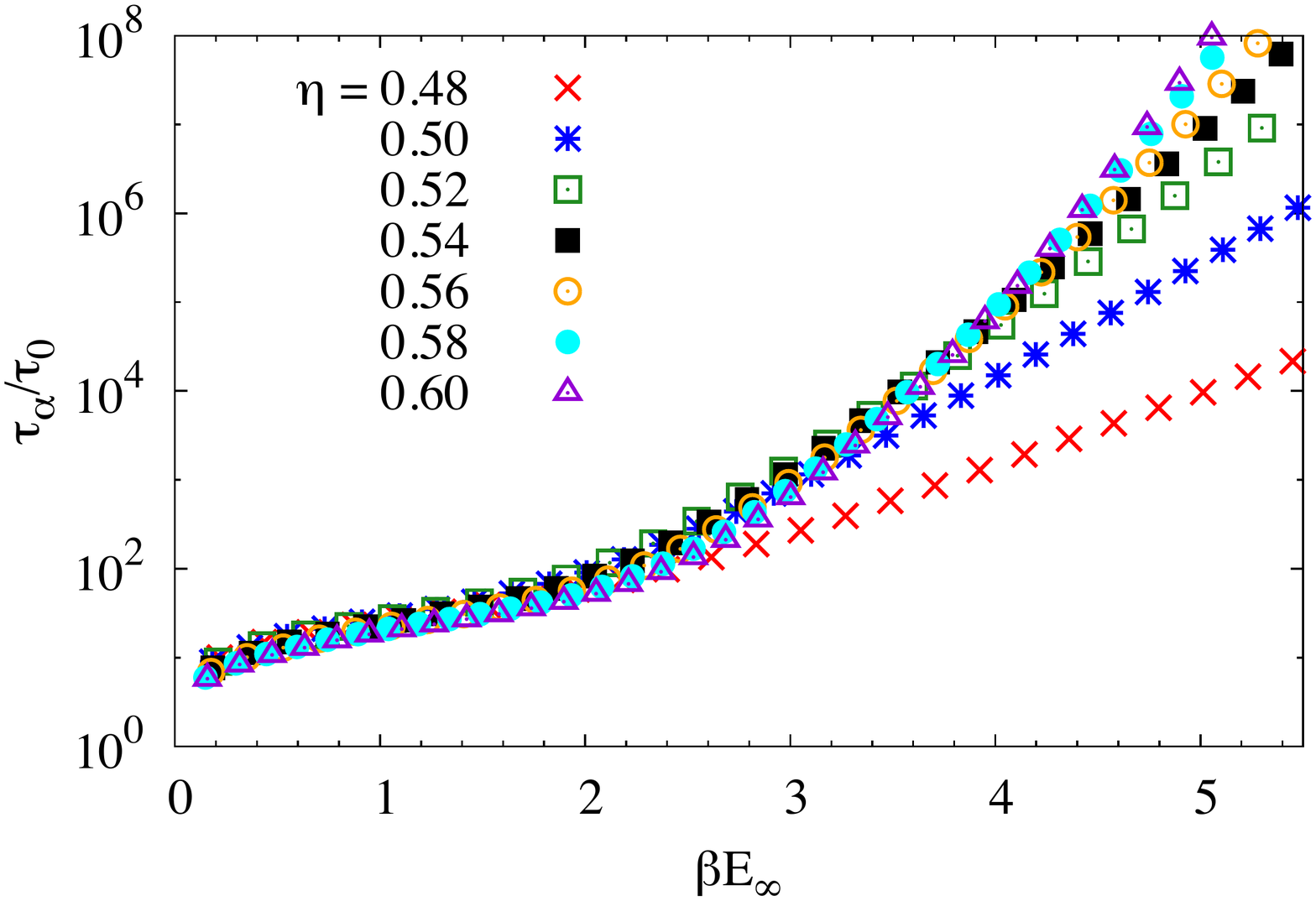}%
	\vspace{-0.5in}
 	\caption{\label{fig:SM2}  Attempt to collapse the non-dimensionalized mean alpha relaxation time for WCA isochoric fluids at various packing fractions. For thermal systems, $\tau_0 \equiv (24 \rho \sigma^2)^{-1} \sqrt{M/\pi k_BT}$, where $M$ is the mass of the particle \cite{JPCL2013MirigianSchweizer, JCP2014MirigianSchweizer1, JCP2014MirigianSchweizer2}.  The temperature is scaled by the high temperature apparent Arrhenius barrier $E_{\infty}$. }
 \end{figure}

Our theory predicts that the isochoric LJ fluid relaxation times can be collapsed using the high temperature Arrhenius barriers (see Fig. 2 in main text), in agreement with simulation  \cite{PRL2009BerthierTarjus,JCP2011BerthierTarjus}. As seen in Fig. \ref{fig:SM2} here, this is not true for the WCA fluid, which is also consistent with simulation \cite{PRL2009BerthierTarjus,JCP2011BerthierTarjus}. Attempts to empirically collapse the WCA alpha times with different energy scales similarly failed.
	
\pagebreak 

\section{E. Inverse Power Law Repulsion Fluid}

 	Pedersen et. al. \cite{PRL2010PedersenSchroder} have shown using isochoric simulation that it is possible to construct a purely repulsive inverse power law (IPL) potential, $U_{IPL}(r) =  A \epsilon (\sigma/r)^n$, that has the same pair correlation function as the LJ fluid if the potential parameters $A$ and $n$ are carefully tuned; for the LJ \textit{mixture} model they studied, one needs $n=15.5$ and $A=1.98$. This tuned IPL fluid exhibits a temperature-dependent alpha relaxation time in excellent agreement with what is found from LJ liquid simulations. Such agreement was argued to be the consequence of an approximate ``hidden scale invariance'', and the strong disagreement of the dynamics of the LJ and WCA fluids arises primarily due to the truncation of the repulsive force for the latter system \cite{PRL2010PedersenSchroder}.

\begin{figure}[b!]
 	\includegraphics[width=0.6\textwidth]{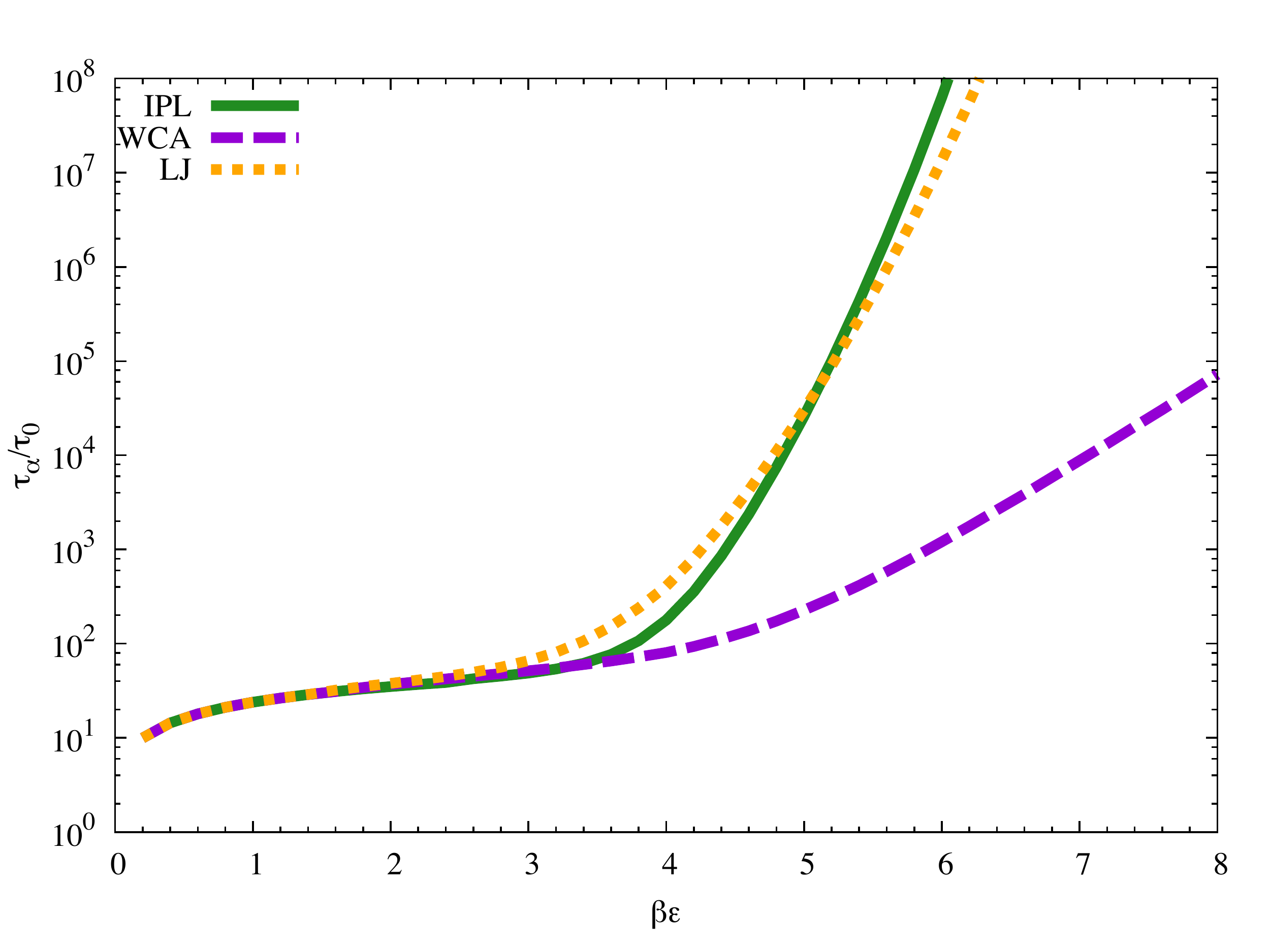}%
	\vspace{-0.2in}
 	\caption{\label{fig:SM3}  The predicted dimensionless alpha relaxation time as a function of the dimensionless inverse temperature for the WCA (dashed purple), LJ (dotted yellow) and IPL (solid green) fluids  at constant $\eta = 0.48$. The IPL potential parameters are $A=0.88$ and $n=15.5$. }
 \end{figure}
 
	We have used PDT to perform a preliminary investigation of this problem. Our system is a one-component fluid, not a binary mixture. But for simplicity, we fix $n=15.5$, and vary $A$ such that the effective temperature-dependent packing fraction of the  IPL, LJ and WCA fluids is nearly identical. Given in the simulation study that the IPL was tuned by hand to reproduce the $g(r)$ of the LJ liquid, we use the same $g(r)$ in our dynamical analysis of the IPL fluids as employed for the WCA and LJ fluids. We then compute the alpha relaxation time at constant volume for this IPL repulsive force fluid using the PDT plus ECNLE theory approach to describe \textit{all} the forces. 

A representative result of our analysis is shown in Figure \ref{fig:SM3}. One sees good agreement between the dynamics of the IPL fluid and the LJ liquid is predicted. This provides additional support for the validity of the PDT idea. Of course, exact agreement is not expected for many reasons: (i) the hidden scale invariance is an approximate idea, (ii) $A=15.5$ is motivated by binary mixture simulations while we study a one-component fluid, (iii) both our dynamical theory and the structural input employed involve statistical mechanical approximations, and (iv) the equilibrium pair structure is computed based on the approximate mapping of the continuous repulsion fluid to an effective hard sphere system. These technical issues will be studied in depth in a future long article.

\section{F. Re-Entrant Glass Melting in Dense Attractive Colloidal Fluids}
	It is well known from experiment  \cite{S2002PhamPuertas}, simulation \cite{JPCM2007Zaccarelli,PRL2002PuertasFuchs,NM2002Sciortino,JPC2005ReichmanRabani}, and microscopic theories based on a projected force vertex (ideal MCT [\cite{JPCM2007Zaccarelli,PRL2002PuertasFuchs,NM2002Sciortino,JPC2005ReichmanRabani} and the local hopping NLE approach \cite{JCP2005Schweizer,JCP2006SaltzmanSchweizer}), that dense hard sphere glasses can dynamically ``melt'' upon the addition of a very short range attraction ($\alpha << d$) of intermediate strength. Simulations  \cite{JPC2005ReichmanRabani}  and colloid experiments \cite{JCP2006KaufmanWeitz} have found that at a fixed high packing fraction, the diffusion constant is a non-monotonic function of the dimensionless attraction strength, $\beta \epsilon$. 
	
Though this problem is well beyond the scope of our Letter, which is focused on viscous liquids, we have performed a demonstration calculation for this system using the PDT idea in the ECNLE framework. Briefly, the cross term between repulsive and attractive forces in Eq. (\ref{eqn:5}) of the main text is retained since both the repulsive and attractive forces are spatially strongly varying and attractive interactions do change local structure, features absent in the LJ liquid. The model pair potential is a hard core repulsion plus exponential attractive tail of contact strength $\epsilon$ and range $\alpha$. Percus-Yevick integral equation theory is used to compute $g(r)$ at a fixed packing fraction $\eta$. 

\begin{figure}[b!]
	\vspace{-0.2in}
 	\includegraphics[width=0.6\textwidth]{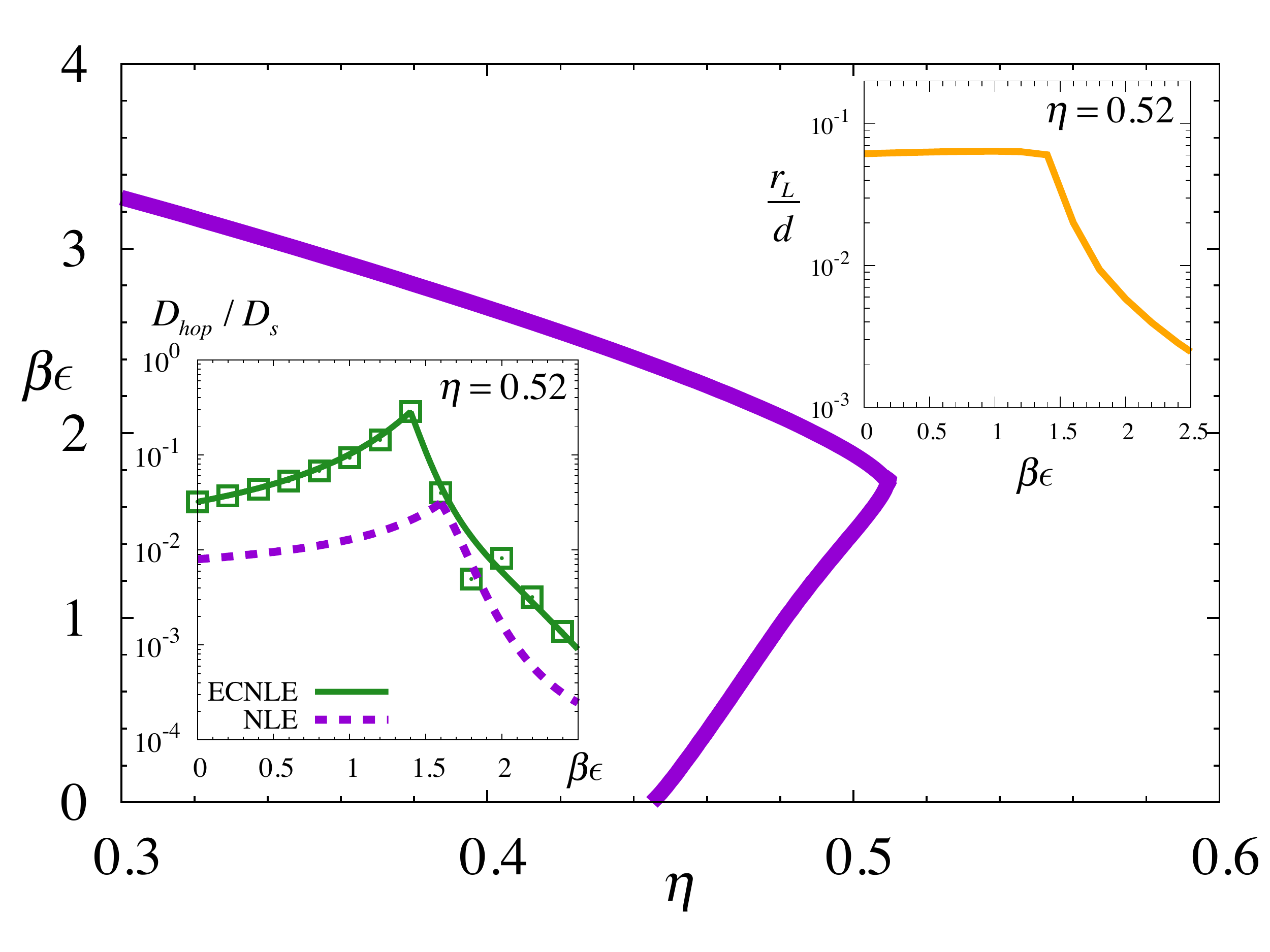}%
	\vspace{-0.2in}
 	\caption{\label{fig:SM4}  Re-entrant phenomena predictions for attractive colloidal systems based on using the PDT force vertex. Main frame - Ideal (NMCT) non-ergodicity boundary. Lower inset - ECNLE and NLE theory results for the dimensionless hopping diffusion constant as a function of reduced attraction strength at a fixed packing fraction of $\eta = 0.52$. Upper inset - Ideal localization length in units of the particle diameter as a function of reduced attraction strength at $\eta = 0.52$.  }
 \end{figure}

Our results based on using the PDT force vertex are shown in Figure \ref{fig:SM4}. The main frame shows the ideal non-ergodicity boundary in the space of packing fraction and attraction strength. In reality, this curve indicates the crossover to a fluid with non-zero barriers and activated dynamics. Its non-monotonic, re-entrant form with the characteristic ``nose'' at $\beta\epsilon \sim 1.7$ is the signature of the ``glass melting'' effect at the ideal kinetic arrest level. The lower left inset shows the dimensionless hopping diffusion constant (in units of the short time diffusivity), $D_{hop}/D_s$ where $D_{hop} \equiv (\Delta r)^2/6\tau_{\alpha, \, hop}$, at a fixed high packing fraction beyond the ``nose''. The combined PDT+ECNLE theory predicts a non-monotonic variation with attraction strength, qualitatively consistent with the findings of simulation \cite{JPC2005ReichmanRabani} and experiment \cite{JCP2006KaufmanWeitz}. Also shown is the prediction of the older NLE theory [4] which captures only the local cage scale hopping physics \cite{JPCL2013MirigianSchweizer,JCP2014MirigianSchweizer1,JCP2014MirigianSchweizer2}.; the behavior is qualitatively the same. The upper right inset shows the corresponding transient localization length (minimum of the dynamic free energy) is roughly constant at a glassÐlike value for relatively weak attractions, and then decreases as the attraction strength grows and physical bonding becomes important. The predicted form of the $\beta \epsilon$-dependence is qualitatively consistent with experiments on colloids \cite{JCP2006KaufmanWeitz} which measured intermediate time plateaus of the single particle mean square displacement. Thus, we conclude that the new PDT+ECNLE approach qualitatively captures the key features of the ``re-entrant glass melting'' phenomenon for dense sticky colloidal fluids. Detailed applications to gel and glass forming particle systems will be pursued in a future long article \cite{DellPhan}. 

\section{G. Equation-of-State for the LJ Liquid}
To model the equation of state of the LJ fluid, we employ the analytic expression:
\begin{align}
\tilde{P} \equiv \frac{\beta P}{\rho} = \frac{1+ \eta_{eff}+\eta_{eff}^2 - \eta_{eff}^3}{(1-\eta_{eff})^3} - C \beta \epsilon\; \eta_{eff}
\label{eqn:SM11} 
\end{align}
The first term in Eq. (\ref{eqn:SM11}) is the classic Carnahan-Starling expression for the (effective) hard sphere fluid \cite{1986HansenMcDonald}. The second term models the role of attractions as a linear in density contribution per the simplest van der Waals picture \cite{1986HansenMcDonald}. The fit parameter is chosen to be $C = 15.7$ in order to best match the LJ model simulation results \cite{HT2003BoltachevBaidakov} at $P=1 \mbox{ atm}$
($\tilde{P} \equiv \beta P/\rho \approx 0$). The inset of figure 3 in the main text shows the solution of Eq. (\ref{eqn:SM11}) compared to the simulation equation of state.  For all higher pressure calculations $C = 15.7$ is held fixed.

\end{document}